%% file: probing-more-deeply-full-o3.tex
\documentclass[aps,prd,twocolumn,amsmath,amssymb,amsfonts,nofootinbib,superscriptaddress,altaffilletter]{revtex4-1}

\usepackage{graphicx}

\usepackage[breaklinks=True]{hyperref}
\hypersetup{
  bookmarksopen=true
}
\usepackage{amssymb}
\usepackage{amsmath}

\usepackage{soul}

\usepackage{color}

\usepackage{supertabular}

\usepackage{dcolumn}
\usepackage{multirow}
\usepackage{xcolor}

\newcommand{\tab}[1]{Table \ref{tab:#1}}

\newcommand{\eqn}[1]{Equation \ref{eqn:#1}}

\newcommand{\sect}[1]{Section \ref{sec:#1}}
\newcommand{\fig}[1]{Figure \ref{fig:#1}}

\def\beq{\begin{equation}}
\def\eeq{\end{equation}}
\def\bea{\begin{eqnarray}}
\def\eea{\end{eqnarray}}

\def\Tcoh{T_{\textrm{\mbox{\tiny{coh}}}}}

\def\sinc{\textrm{sinc}}

\newif\ifshowfigs
\showfigstrue

\newcommand{\Fstat}{$\mathcal{F}$-statistic}

\newcommand\T{\rule{0pt}{2.6ex}}       
\newcommand\B{\rule[-1.2ex]{0pt}{0pt}} 

\def\pf{{\footnotesize\textsc{PowerFlux}}}
\def\pyfstat{{\footnotesize\textsc{PyFstat}}}
\def\fh{{\it FrequencyHough}}
\def\sh{{\it SkyHough}}

\def\sci#1#2{#1\times10^{#2}}

\def\RAJ{\textrm{RA}_{\textrm J2000}}
\def\DECJ{\textrm{DEC}_{\textrm J2000}}

\begin{document}

\title{
Probing More Deeply in an All-Sky Search for \\ Continuous Gravitational Waves in the LIGO O3 Data Set
}

\let\mymaketitle\maketitle
\let\myauthor\author
\let\myaffiliation\affiliation
\author{Aashish Tripathee}
\affiliation{University of Michigan Physics Department, Ann Arbor, MI 48109, USA}
\author{Keith Riles}
\affiliation{University of Michigan Physics Department, Ann Arbor, MI 48109, USA}

\date[\relax]{compiled \today}

\noaffiliation

\begin{abstract}
  We report results from an all-sky search of the LIGO data from the third LIGO-Virgo-KAGRA run (O3) for continuous gravitational waves from isolated neutron stars in the frequency band [30, 150] Hz and spindown range of $[\sci{-1}{-8}, \sci{+1}{-9}]$ Hz/s.
  This search builds upon a previous analysis of the first half of the O3 data using the same \pf\ pipeline.
  We search more deeply here by using the full O3 data and by using loose coherence in the initial stage with fully coherent combination of LIGO Hanford (H1) and LIGO Livingston (L1) data,
  while limiting the frequency band searched and excluding narrow, highly disturbed spectral bands.
  We detect no signal and set strict frequentist upper limits on circularly polarized and on linearly polarized wave amplitudes,
  in addition to estimating population-averaged upper limits.
  The lowest upper limit obtained for circular polarization is $\sim\sci{4.5}{-26}$,
  and the lowest linear polarization limit is $\sim \sci{1.3}{-25}$ (both near 144 Hz).
  The lowest estimated population-averaged upper limit is $\sim\sci{1.0}{-25}$.
  In the frequency band and spindown range searched here, these limits improve upon the O3a \pf\ search by a median factor of $\sim 1.4$ and upon the best previous limits obtained
  for the full O3 data by a median factor of $\sim 1.1$.
\end{abstract}

\maketitle

\section{Introduction}
\label{sec:introduction}

LIGO~\cite{bib:AdvancedLIGO}, VIRGO~\cite{bib:VIRGO}, and KAGRA~\cite{bib:KAGRA1,bib:KAGRA2,bib:KAGRA3} are interferometer-based experiments that search for gravitational waves. Nearly 100 transient gravitational waves have been detected from black hole (BH) mergers, neutron stars (NS) mergers, and BH-NS mergers since the first detection in 2015~\cite{bib:FirstGWDetection,bib:GWTC1,bib:GWTC2,bib:GWTC3,bib:NSBH}. An as yet undetected type of gravitational waves is continuous, long-lived, and nearly monochromatic. Such radiation is necessarily weaker than that detected already from transient mergers and requires long-period integration for detection in the presence of interferometer noise~\cite{bib:LRRReview}.

In this work we focus on one of the potential sources for continuous gravitational waves -- rapidly spinning isolated neutron stars with non-axisymmetry. While previous all-sky searches in the O3 data~\cite{bib:cwallskyO3aPowerFlux,bib:cwallskyO3FourPipelines}, including the O3a \pf\ analysis, analyzed a broad band of approximately 20-2000 Hz, this work focuses on young rapidly spinning neutron stars in the Milky Way with gravitational wave frequencies of 30-150 Hz. The non-axisymmetry in such stars could stem from different sources including crustal deformation, buried magnetic field energy, and excitation of r-modes~\cite{bib:Lasky_review,bib:GandG_review,bib:BejgerReview}.

This search uses the venerable \pf\ pipeline~\cite{bib:PowerFlux1,bib:PowerFlux2,bib:PowerFluxPol,bib:cwallskyearlyS5,bib:cwallskyS5,bib:cwallskyS6,bib:cwallskyO1paper1,bib:cwallskyO1paper2} with loose coherence~\cite{bib:loosecoherence,bib:cwallskyS5,bib:LooseCoherenceMediumScale,bib:LooseCoherenceWellModeledSignals} to perform an all-sky search in the spindown range $[\sci{-1}{-8}, \sci{+1}{-9}]$ Hz/s. Our search resulted in $\approx \sci{3.4}{6}$ outliers\footnote{This number is before clustering which reduced the number to $\approx \sci{2.9}{6}$.} in the first stage
after excluding narrow, highly disturbed bands, a number that dropped to $\approx \sci{1.2}{4}$ by the final stage of the hierarchical search. These outliers were then followed up on with the Markov Chain Monte Carlo (MCMC) based \pyfstat\ framework~\cite{bib:pyFstat,bib:TenorioEtal}. All but 10 of the surviving outliers can be attributed to hardware injections (simulated CW signals imposed upon the interferometer mirror motion) or to known spectral artifacts. Those 10 outliers, however, have Bayes factors obtained from \pyfstat\ much less than the determined threshold expected for a plausible signal. As this search failed to detect a true signal, we set upper limits on strain amplitudes as a function of frequency, excluding excised bands.

This article is organized as follows: \sect{dataset} describes the data set we used. \sect{analysis_method} discusses the analysis method with \pf\ and loose-coherence in addition to the data cleaning that was performed to suppress instrumental artifacts. \sect{results} shows the upper limits and outliers results obtained. \sect{conclusions} concludes the article with a discussion of the results and prospects for future searches including the recently started O4 search.

\section{Data Sets Used}
\label{sec:dataset}

This analysis uses data from the O3 run of Advanced LIGO. Advanced LIGO consists of two interferometers -- one at Hanford, Washington (H1) and the other at Livingston, Louisiana (L1), separated by a 3000 km baseline~\cite{bib:AdvancedLIGO}. The O3 run took place between April 1, 2019 and March 27, 2020~\cite{bib:O3OpenData}. The first part of the run (O3a) and the second part (O3b) were separated by a commissioning break in October 2019~\cite{bib:O3OpenData}. This search builds upon the O3a \pf\ analysis~\cite{bib:cwallskyO3aPowerFlux} but uses the $\sim$11 months of the full O3 dataset.

Before the analysis was performed, known spectral artifacts (``lines'')~\cite{bib:GoetzEtalO3linelists} were excised and replaced with random Gaussian noise (``cleaned''). Additionally, loud single-detector artifacts of an unknown source were also cleaned~\cite{bib:GoetzEtalO3linelists}. Details of the cleaning are discussed below in \sect{data_cleaning}.

Another artifact affecting the data were loud, relatively frequent glitches -- short, high amplitude instrumental transients~\cite{bib:SelfGating1,bib:SelfGating2,bib:O2O3DetChar}. While also present in previous runs, they were much louder and more frequent in the O3 run~\cite{bib:cwallskyO3aPowerFlux,bib:O2O3DetChar}. The effect of glitching was particularly troublesome for frequencies below 500 Hz. We started from ``self-gated'' data in which a narrow inverse-Tukey-window zeroing gate was applied to the time-domain data, details of which can be found in the technical document~\cite{bib:SelfGating1}. Gating applied here leads to livetime losses of about 3\% and 12\%

of the H1 and L1 observation times. The calibration uncertainties of this C01 gated data set are estimated to be $<$ 7\%\ in magnitude and $<$ 4 deg in phase for O3a and $<$ 11\%\ in magnitude and $<$ 9 deg in phase for O3b (68\%\ confidence interval)~\cite{bib:O3aCalibPaper,bib:O3bCalibPaper}.

\section{Analysis Method}
\label{sec:analysis_method}

This analysis uses the \pf\ pipeline to search for continuous gravitational waves. The strain data is divided into 7200 s segments for which discrete Fourier transforms are computed (``short'' Fourier Transforms -- SFTs), using Hann windowing
with 50\%\ overlap. For a large number of templates that include corrections for spindown evolution and Doppler modulation, power is summed up over all the SFTs while using an inverse-noise-weighting to disfavor SFTs with high noise levels~\cite{bib:cwallskyS4}. For each 1/16 Hz search sub-band, the power is maximized across the sky and across spindown steps to produce 95\% frequentist upper limits~\cite{bib:cwallskyS4,bib:cwallskyO1paper1,bib:cwallskyO1paper2}. The first stage of the hierarchical search, called Stage 0, determines upper limits and yields a set of outliers defined by a signal-to-noise ratio (SNR) greater than 7. These outliers undergo two stages of loose coherence followup, applying tighter constraints and requiring increased SNR, after which persistent outliers are manually examined via strain histograms~\cite{bib:cwallskyO3aPowerFlux} to identify instrumental artifacts~\cite{bib:StrainHistogram}. Surviving outliers are then followed up by the \pyfstat\ pipeline which determines a Bayes factor of preference for signal over background for each outlier~\cite{bib:pyFstat,bib:TenorioEtal}.

\subsection{Signal Model}

Our analysis assumes a model in which a spinning neutron star with a time-varying quadrupole moment produces circularly polarized waves along the direction of the spin-axis and linearly polarized waves in the perpendicular directions~\cite{bib:cwallskyS4,bib:cwallskyO1paper1,bib:LRRReview} with a frequency twice that of the star's spin frequency. The polarization of the detected wave then depends on the inclination angle between the spin-axis and the line-of-sight to the Earth, $\iota$~\cite{bib:LRRReview}. Extreme cases are $\iota = 0, \pi$ and $\iota = \pi / 2$. The former corresponds to circularly polarized waves which is the best-case scenario maximizing incident signal power. The latter corresponds to linearly polarized waves which is the worst-case scenario.

Specifically, the signal model we assume is:

\begin{align}
h(t) &= h_0 \Big\{ F_+(t, \alpha_0, \delta_0, \psi) \frac{1 + \cos{\iota}^2}{2} \cos{\left[ \Phi(t) \right]} \nonumber \\
&+ F_\times (t, \alpha_0, \delta_0, \psi) \cos{\iota} \sin{\left[ \Phi(t) \right]}\Big\} \;,
\label{eqn:signal_model}
\end{align}
where $h_0$ is the intrinsic strain amplitude, $\Phi(t)$ is the signal phase, the $F_+$ and $F_\times$ are detector responses to signals with plus and cross polarizations, respectively~\cite{bib:cwallskyS4,bib:cwallskyO1paper1,bib:LRRReview}. 

In the loose coherence stages of the search, we assume a second-order Taylor series
model~\footnote{The final \pyfstat\ search stage that can include much longer coherence times also accounts for higher frequency derivatives, as needed.}
for the phase evolution in which we include only terms up to quadratic
in offsets from a reference time $\tau$:

\beq
\Phi(\tau) = \Phi_0 + 2 \pi \left\{ f ( \tau - \tau_\textrm{ref}) + {\frac{1}{2}}\dot{f} (\tau - \tau_\textrm{ref})^2 \right\} \;,
\eeq
where $\Phi_0$ is the initial phase and $f, \dot{f}$ are the frequency and spindowns at the chosen reference time $\tau_\textrm{ref}$:

\beq
\tau_\textrm{ref}(t) = t + \frac{\vec{r}(t) \cdot \vec{n}}{c} + \Delta_{E_\odot} - \Delta_{S_\odot} \;,
\eeq
where $\vec{r}(t)$ is the position vector of the detector in the Solar System Barycenter (SSB) frame, $\Delta_{E_\odot}$, and $\Delta_{S_\odot}$ are the relativistic Einstein and Shapiro time delays respectively~\cite{bib:cwallskyO3FourPipelines}. $\vec{n}$ is the vector pointing from the detector toward the source, $\vec{n} = (\cos{\alpha} \cos{\delta}, \sin{\alpha} \cos{\delta}, \sin{\delta})$.

In \eqn{signal_model}, $h_0$ is the intrinsic strain amplitude and is a function of the ellipticity:

\begin{align}
h_0 &= \frac{ 4 \pi^2 G}{c^4} \frac{\epsilon I_{zz} f^2}{d}  \nonumber\\
\approx &1.06 \times 10^{-26} \left( \frac{\epsilon}{10^{-6}} \right) \left(\frac{I_{zz}}{10^{38}~kg~m^2} \right) \nonumber\\
&\left(\frac{f}{100~\mathrm{Hz}} \right)^2 \; \left( \frac{1~\textrm{kpc}}{d} \right)\;,
\label{eqn:strain_relation}
\end{align}
where $\epsilon$ is the equatorial ellipticity of the star, $\frac{I_{xx} - I_{yy}}{I_{zz}}$, which is a dimensionless measure of the non-axisymmetry of the star with $I_{xx,yy,zz}$ being the star's principal moments of inertia (spin axis along $z$),
  $d$ is the distance from the source to the detector, and $f$ is the gravitational wave frequency~\cite{bib:cwallskyS4,bib:LRRReview}. If we assume that the loss of rotational energy of the source is solely through gravitational wave emission, we can define the spindown limit on the maximum detectable strain, $h_{\mathrm{sd}}$ \cite{2007PhRvD..76h2001A}:

\begin{align}
h_{\mathrm{sd}} = &(2.5 \times 10^{-25}) \left( \frac{1~\mathrm{kpc}}{d} \right) \cdot \notag \\&\sqrt{ \left( \frac{1~\mathrm{kHz}}{f} \right) \left( \frac{- \dot{f}}{10^{-10}~\mathrm{Hz/s}} \right) \left( \frac{I_{zz}}{10^{38}~kg~m^2} \right) } \;.
\label{eqn:h_sd_equation}
\end{align}

\subsection{Parameter Space Analyzed}

\begin{figure*}[htbp]
  \begin{center}
    \ifshowfigs
    \includegraphics[width=0.9\textwidth]{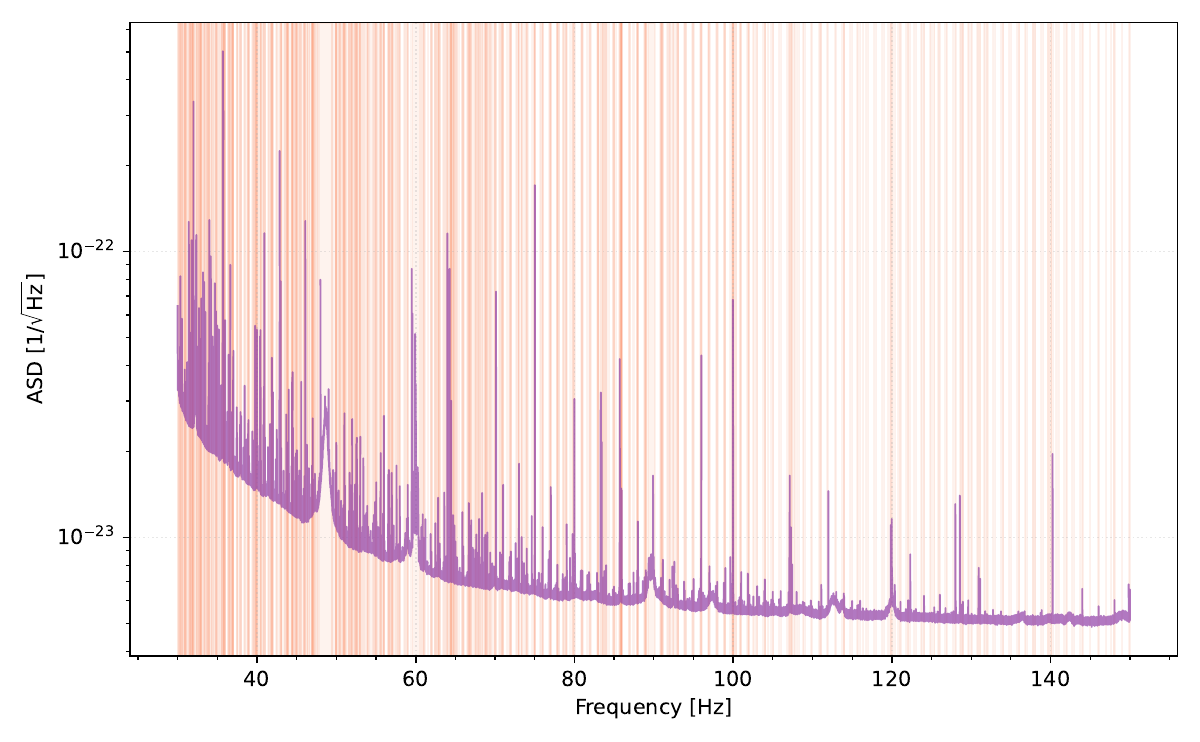}
    \includegraphics[width=0.9\textwidth]{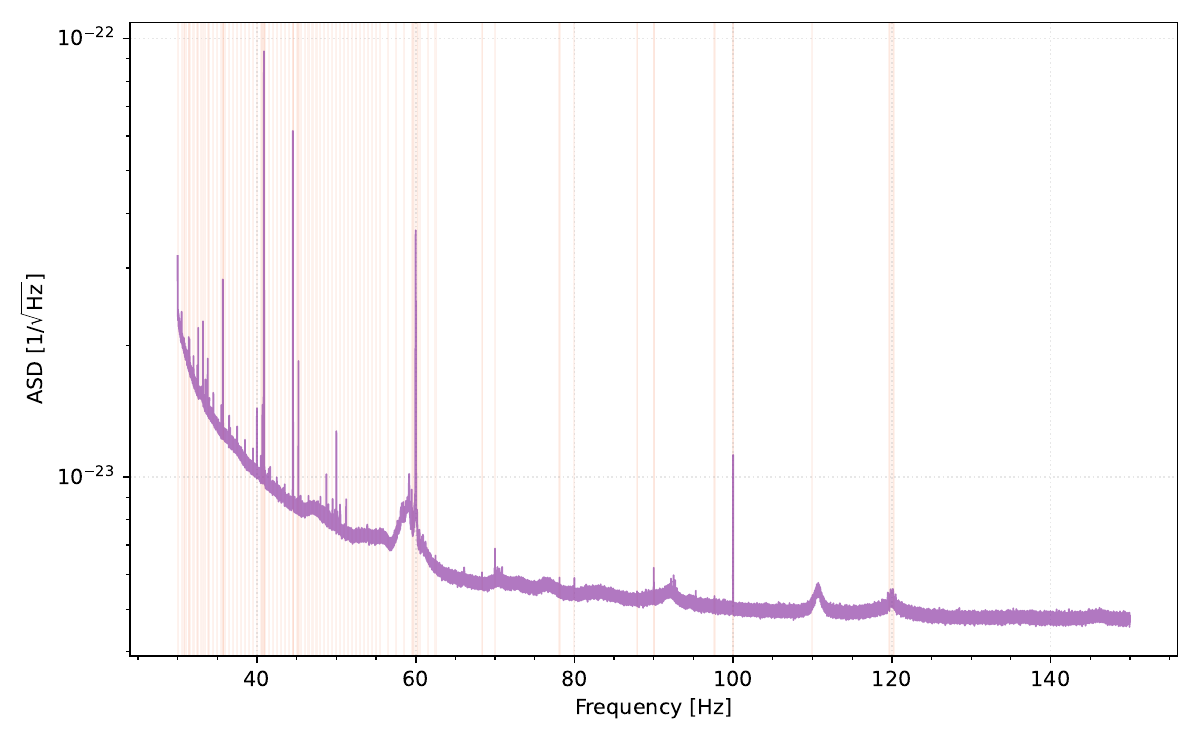}
    \fi
    \caption[Cleaned bands]{Cleaned sub-bands for H1 (top) and L1 (bottom). The purple lines show the respective amplitude spectral density (ASDs), while the shaded vertical bars show regions where the data was replaced with randomly generated Gaussian noise. In most cases, the thickness of the vertical bar shown is larger than the width of the corresponding cleaned subband. The fractions of bandwidth cleaned are 6.3\%\ and 1.2\%\ for H1 and L1 data, respectively.}
   \label{fig:excluded_bins}
  \end{center}
\end{figure*}

\begin{table*}[htbp]
\begin{center}
\begin{tabular}{llccccc}\hline
\T\B Stage & Instrument sum & {Phase coherence $\delta$} & \multicolumn{1}{c}{Spin-down step} & \multicolumn{1}{c}{Sky refinement} & \multicolumn{1}{c}{Frequency refinement} & \multicolumn{1}{c}{SNR increase} \\
 & & \multicolumn{1}{c}{rad} & \multicolumn{1}{c}{Hz/s} & $\alpha\times\delta_0$  &  & \multicolumn{1}{c}{\%}\\
\hline \hline
\multicolumn{7}{c}{\T\B 30-60\,Hz frequency range} \\
  0 & Initial/upper limit coherent & $\pi/2$      & $\sci{5.0}{-11}$ & $1/2 \times 1/2$   & $1/4$ & -- \\
  1 & Coherent                        & $\pi/4$ & $\sci{5.0}{-12}$   & $1/4\times1/4$ & $1/8$ & 10 \\
  2 & Coherent                          & $\pi/8$ & $\sci{1.0}{-12}$   & $1/8\times1/8$ & $1/16$ & 20  \\
 \hline
\multicolumn{7}{c}{\T\B 60-90\,Hz frequency range} \\
  0 & Initial/upper limit coherent & $\pi/2$      & $\sci{3.0}{-11}$ & $1/2 \times 1/2$   & $1/4$ & -- \\
  1 & Coherent                        & $\pi/4$ & $\sci{5.0}{-12}$ & $1/4\times1/4$ & $1/8$ & 10 \\
  2 & Coherent                          & $\pi/8$ & $\sci{1.0}{-12}$   & $1/8\times1/8$ & $1/16$ & 15  \\
 \hline
  \multicolumn{7}{c}{\T\B 90-120\,Hz frequency range} \\
  0 & Initial/upper limit coherent & $\pi/2$      & $\sci{3.0}{-11}$  & $1/2 \times 1/2$   & $1/4$ & -- \\
  1 & Coherent                        & $\pi/4$ & $\sci{1.0}{-11}$  & $1/4\times1/4$ & $1/8$ & 10 \\
  2 & Coherent                          & $\pi/8$ & $\sci{5.0}{-12}$   & $1/8\times1/8$ & $1/16$ & 25  \\
 \hline
 \multicolumn{7}{c}{\T\B 120-150\,Hz frequency range} \\
  0 & Initial/upper limit coherent & $\pi/2$      & $\sci{3.0}{-11}$  & $1/2 \times 1/2$   & $1/4$ & -- \\
  1 & Coherent                        & $\pi/4$ & $\sci{5.0}{-12}$  & $1/4\times1/4$ & $1/8$ & 10 \\
  2 & Coherent                          & $\pi/8$ & $\sci{1.0}{-12}$   & $1/8\times1/8$ & $1/16$ & 25  \\
 \hline
\end{tabular}
\caption[Outlier followup parameters]{
Outlier followup parameters. All stages use loose coherence for demodulation and sum Hanford and Livingston data coherently.}
\label{tab:followup_parameters}
\end{center}
\end{table*}

In this work, we search more deeply than in the 2021 O3a all sky LVK search~\cite{bib:cwallskyO3aPowerFlux}, as discussed in Section~\ref{sec:methodology}, but limit the search band to lower frequencies (30--150 Hz) in order partly to make the total computational cost manageable
and partly to focus on young neutron stars with lower spin frequencies (15--75 Hz). Excluding the 20--30 Hz band searched in the O3a analysis was a pragmatic choice, given the scarcity of clean spectral subbands in that range and a highly elevated
noise floor in those subbands relative to that at higher frequencies.

We maintain the same spindown range of $\left[ - \sci{1}{-8}, + \sci{1}{+9} \right]$ Hz/s as in the O3a analysis~\cite{bib:cwallskyO3aPowerFlux}. Most isolated sources of continuous gravitational waves are expected to have a negative frequency derivative as they lose energy through the emission of gravitational waves~\cite{bib:LRRReview,bib:cwallskyO3aPowerFlux}. A small number of isolated pulsars in globular clusters are known to have slight apparent spin-ups however. These are believed to arise from their acceleration in the Earth's direction~\cite{bib:LRRReview,bib:cwallskyO3aPowerFlux}. Other reasons could include a strong slowly-varying Doppler shift from a long-period orbit with a binary companion~\cite{bib:LRRReview,bib:cwallskyO3aPowerFlux}. Exotic sources with spin-ups could include superradiance from a boson cloud in the vicinity of an isolated black hole~\cite{bib:axiverseArvanitaki2}.
Retaining a large search range in negative spin derivative allows probing sources at larger distances $d$
as constrained by consistency among detectable $h_\mathrm{sd}$ and $d$ in
Equation~\ref{eqn:h_sd_equation} and implied rotational energy loss.

\subsection{Data Cleaning}
\label{sec:data_cleaning}

The low-frequency spectrum, particularly that of Hanford H1 data, suffered from a large number of spectral artifacts. This led to an overwhelming number of outliers during the O3a analysis which made outlier followup difficult and computationally costly. Having performed an ``eyes wide open'' search in the O3a data, we adopt a more pragmatic
approach here, excising heavily polluted subbands, replacing the SFT coefficients by Gaussian random noise with
target strain power based on neighboring frequency bands~\cite{bib:lal}. This approach avoids inundation by
outliers in bands for which true signal detection likelihood is {\it a priori} low.

We perform line-cleaning in two steps. First, we clean out  lines that have a known instrumental source using the list presented in~\cite{bib:GoetzEtalO3linelists}. One approach to determining that the source of a line is unlikely to be astrophysical in origin involves looking for excess correlation between the strain channel and an environmental channel that is expected to be decoupled from the strain channel \cite{bib:DavisEtalDetchar}. More details can be found in \cite{bib:linesO1O2, bib:DavisEtalDetchar}.

A large number of single-detector lines of unknown source also affect the
bands, however, some of which we clean in the second step. We remove only those
  lines clearly inconsistent with an astrophysical source, while accounting for the differing H1 and L1 noise floors.
  Specifically for a strain power spectral noise line peak in detector A of median-filter applied maximum value PSD$^A_{\rm max}$
  and a visibly smooth (non-peaked) corresponding spectral range in detector B,
  we deem the peak in A to be non-astrophysical if median-filter applied PSD$^A_{\rm max}$ is greater than
  $2.5$ times the counterpart noise level PSD$^B$ (also median-filter applied). We use this criterion to mark regions to clean, which we then refine slightly manually. Injection studies confirmed the safety of our vetoes, given that sky locations that may lie near an antenna pattern node for detector B at one moment, while retaining
  sensitivy in detector A will not sustain this imbalance when averaged over sidereal time. \fig{excluded_bins} shows the bands we cleaned. Our data cleaning led to 6.3\% of the highly contaminated H1 band and 1.2\% of the L1 band being cleaned.

\subsection{Methodology}
\label{sec:methodology}

\begin{figure*}[htbp]
  \begin{center}
    \ifshowfigs
    \includegraphics[width=3.3in]{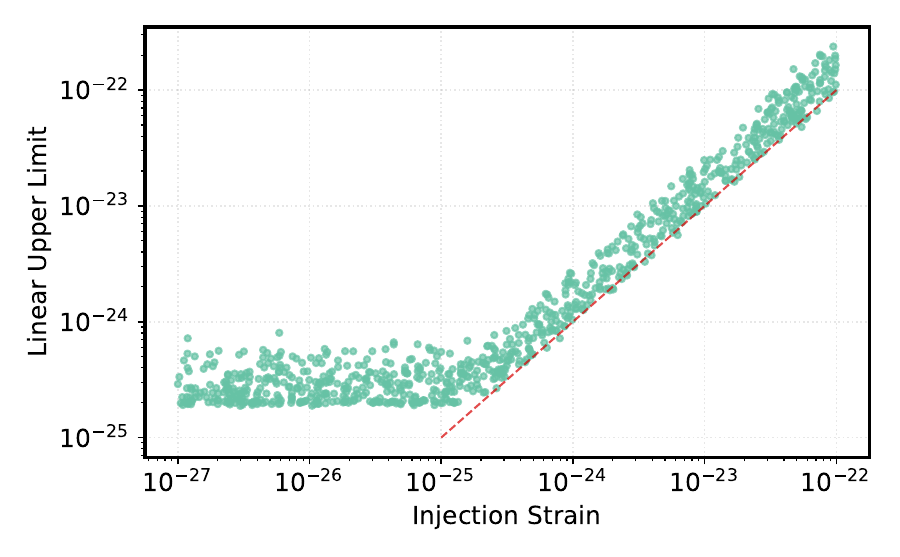}
    \includegraphics[width=3.3in]{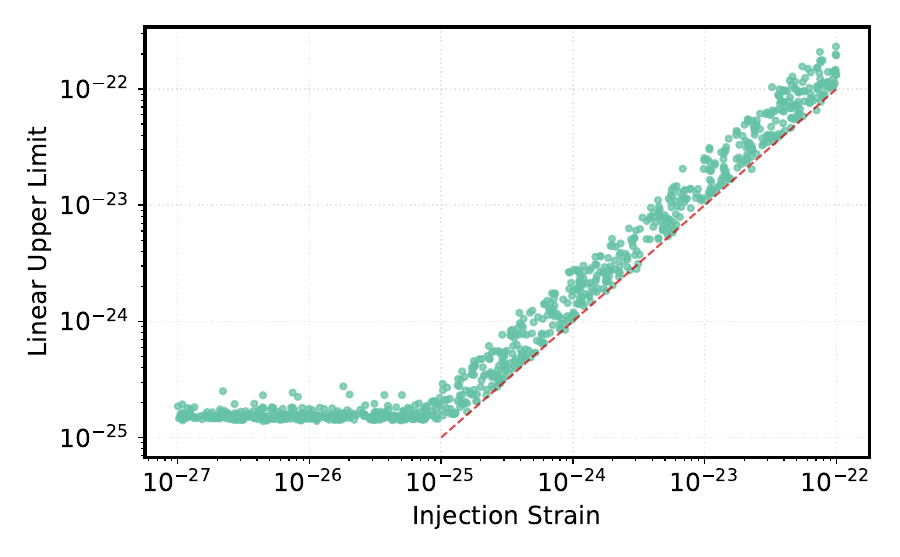}
    \includegraphics[width=3.3in]{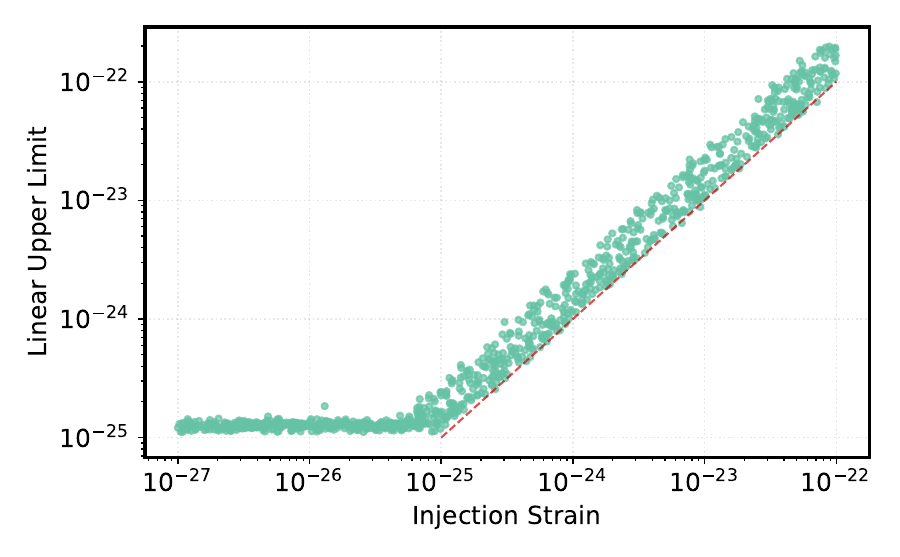}
    \includegraphics[width=3.3in]{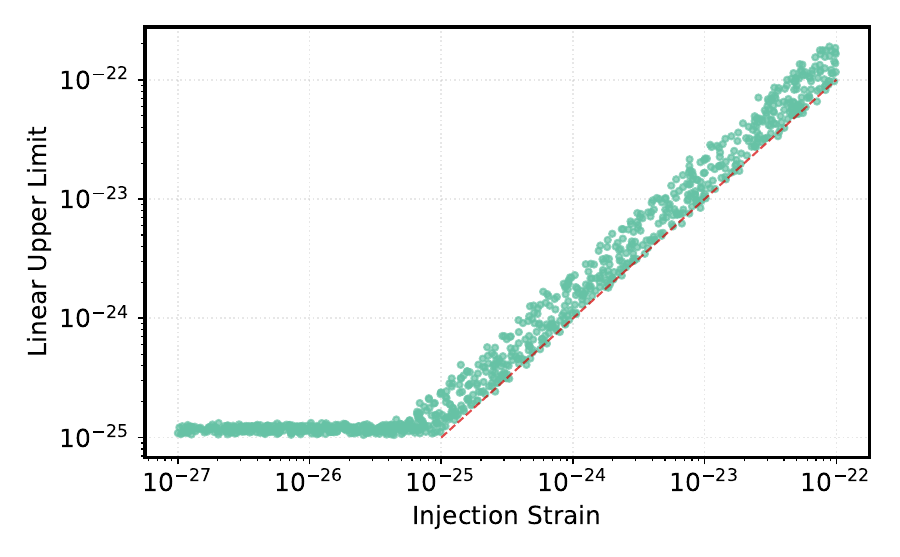}
    \fi
    \caption[Upper limits validation]{Upper limits validation for the four major 30 Hz sub-bands -- 30-60 Hz, 60-90 Hz, 90-120 Hz, and 120-150 Hz from top-left to bottom-right. Each point represents a unique injection with randomly generated parameter including polarization for which the produced upper limit is plotted against the injected strain. The red diagonal lines define the equality of upper limit with the injected strain.}
   \label{fig:ul_vs_strain}
  \end{center}
\end{figure*}

We follow a hierarchical search strategy using the \pf\ pipeline~\cite{bib:PowerFlux1,bib:PowerFlux2,bib:PowerFluxPol,bib:cwallskyearlyS5,bib:cwallskyS5,bib:cwallskyS6,bib:cwallskyO1paper1,bib:cwallskyO1paper2} with loose coherence~\cite{bib:loosecoherence,bib:cwallskyS5,bib:LooseCoherenceMediumScale,bib:LooseCoherenceWellModeledSignals}. While the first stage (called Stage 0) establishes upper limits and produces first round of outliers, subsequent stages follow up on those outliers with increasing levels of constraints on signal coherence and parameter reconstruction. We steadily increase the effective coherence times by halving the phase mismatch parameter, $\delta$ between each successive stages, thereby increasing the sensitivity at each new stage~\cite{bib:loosecoherence,bib:cwallskyS5}. Additionally, we reduce the step-size for frequency, spindown, and sky location after each stage. As each successive stage improves sensitivity, we also require the SNR to increase while going from one stage to the next one. The exact factor of increase we require depends on the band and was empirically determined via software signal injections into the O3 data. \tab{followup_parameters} shows the values of phase-mismatch parameter $\delta$, step-sizes, and SNR increase factors we require for each of the three stages. Outliers that survive all stages are then manually investigated through the use of strain histograms to look for contamination from known instrumental artifacts or single-interferometer lines with unknown origins. Outliers that survive all of these requirements are then followed up with the \pyfstat\ pipeline~\cite{bib:pyFstat}, as in the O3a analysis~\cite{bib:cwallskyO3aPowerFlux}.

\begin{figure*}[htbp]
  \begin{center}
    \ifshowfigs
    \includegraphics[width=3.3in]{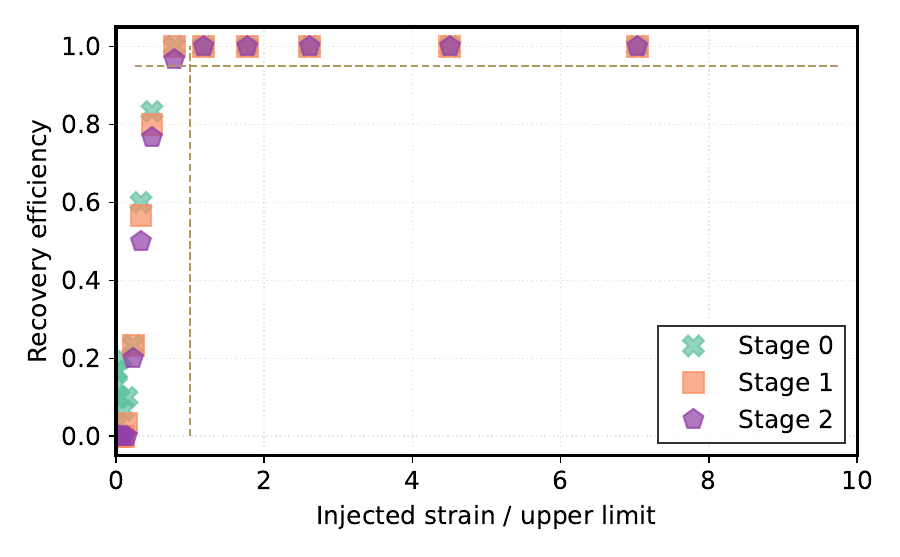}
    \includegraphics[width=3.3in]{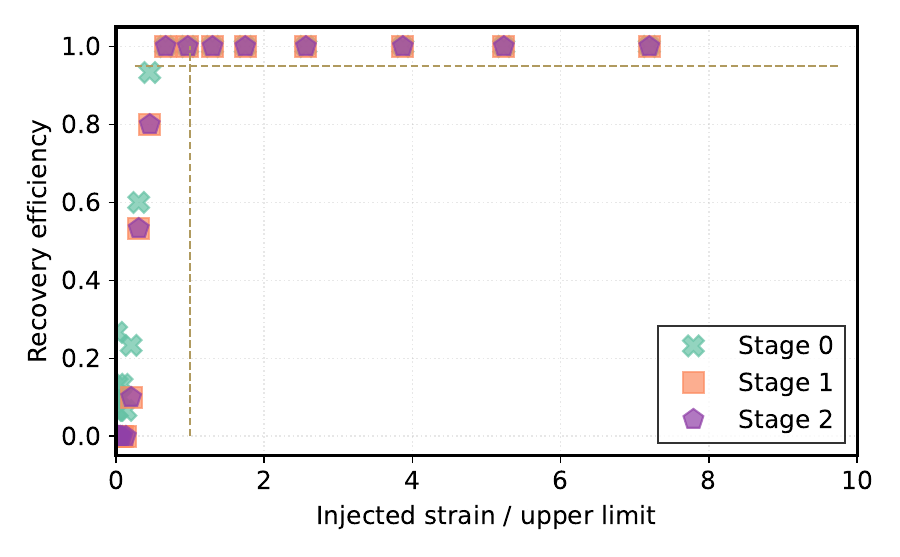}
    \includegraphics[width=3.3in]{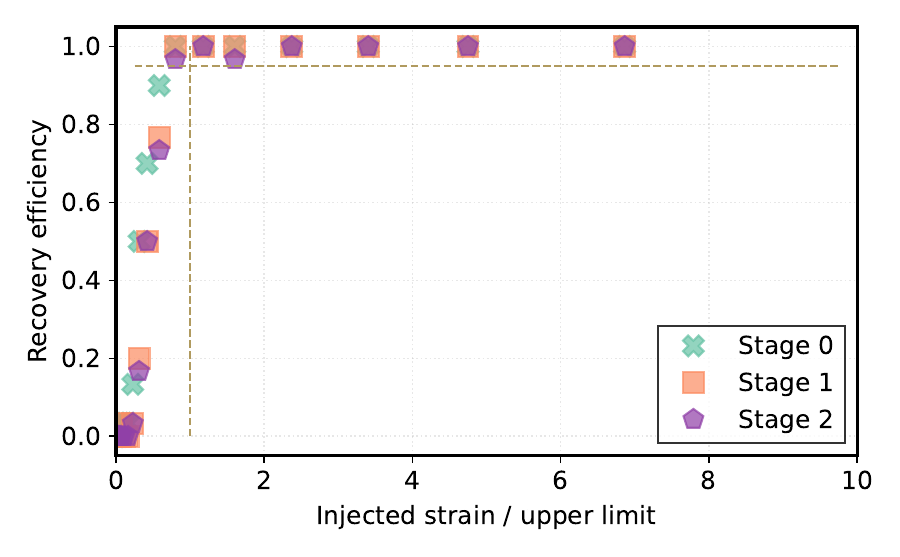}
    \includegraphics[width=3.3in]{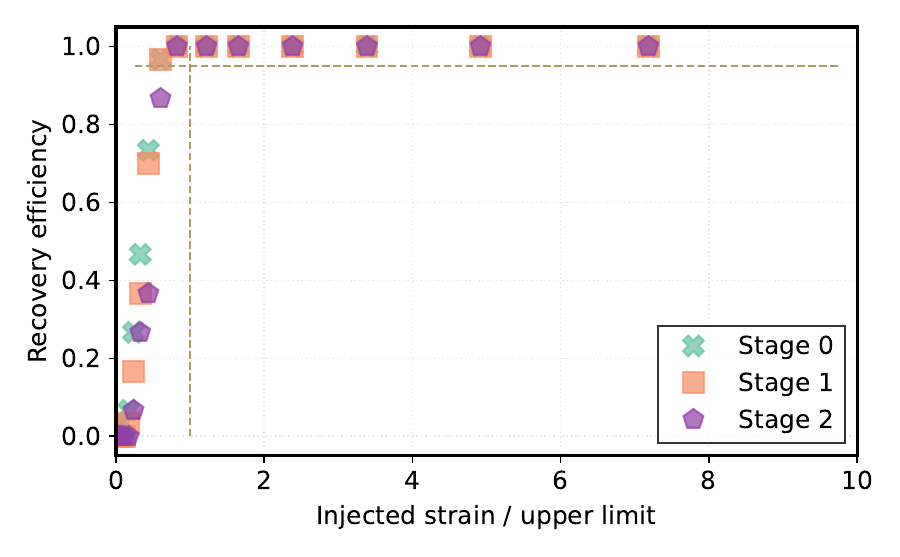}
    \fi
    \caption[Injection recovery]{Injection (software simulations) recovery efficiencies
      in the 30-60\,Hz, 60-90\,Hz, 90-120\,Hz, and 120-150\,Hz frequency bands [top-left to bottom-right]. The horizontal axis shows the relative upper limit -- the ratio of the injected strain to the 95\%\ CL upper limit in the corresponding band without any injection. The vertical axis shows the fraction of surviving injections. The horizontal dashed lines correspond to a 95\%\ recovery fraction while the vertical dashed lines represent a relative strain of 1 where the injected strain is equal to the 95\%\ upper limit.}
   \label{fig:injection_recovery}
  \end{center}
\end{figure*}

\pf\ iterates through all SFTs and computes the power for each template through a bilinear form of the (SFT) input matrix $\{ a_{t, f} \}$, indexed by the SFT time $t$ and template frequency $f$:
\bea
P[f] = \sum_{t_1, t_2, D_i, D_j} a_{t_1, f + \Delta f (t_1)}^{(D_i)} \; a_{t_2, f + \Delta f (t_2)}^{(D_j)^*} \; K_{t_1, t_2, f}^{D_i D_j} \;,
\eea
where $\Delta f$ is the detector frequency shift arising from Doppler shift due to the Earth's motion and the neutron star's spindown~\cite{bib:cwallskyO1paper1,bib:cwallskyO1paper2}. The factor $K_{t_1, t_2, f}^{D_i D_j}$ includes a Lanczos kernel along with contributions from time-dependent SFT weights, antenna patterns, signal polarization, and relative terms for the detectors $D_{i,j}$~\cite{bib:cwallskyS5,bib:loosecoherence,bib:cwallskyO1paper1,bib:cwallskyO1paper2}. The phase-mismatch parameter $\delta$ in the Lanczos kernel portion of $K_{t_1, t_2, f}^{D_i D_j}$:
\beq
\widetilde{K}_{\delta}(t_1, t_2) = \sinc{ \left[ \frac{\delta~(t_1 - t_2)}{\Tcoh} \right] }~\sinc{ \left[ \frac{\delta~(t_1 - t_2)}{3~\Tcoh} \right] } \;,
\eeq
where $t_1, t_2$ are the mid-times of the two SFTs in the barycenter frame, and $T_{\mathrm{coh}} = 7200~s$, governs the degree of signal phase drift we tolerate~\cite{bib:loosecoherence,bib:cwallskyS5}. Loose coherence helps discard templates with non-physical phase evolution and hence, distinguish between signals that are possibly astrophysical in origin from instrumental artifacts~\cite{bib:loosecoherence}. Steadily decreasing the value of $\delta$ used reduces the amount of phase drift permitted in each successive stage, hence imposing tighter constraints on the templates that merit followup with another stage of still tighter constraints~\cite{bib:loosecoherence,bib:cwallskyS5}.

The choice of the Lanczos kernel (or a sinc filter) for loose coherence essentially makes it equivalent to performing a coherent search with a low-pass filter that attenuates signals with rapidly varying phases \cite{2010CQGra..27t5017D,tripathee2022all}. The cutoff frequency is determined by the tunable phase mismatch parameter $\delta$, and hence lowering its value in follow-up stages  of the search demand progressively higher coherence between SFTs, thereby disfavoring templates with non-physical phase evolutions \cite{2010CQGra..27t5017D,2012PhRvD..85f2003D}.

As our goal is to improve upon the earlier O3a \pf\ search~\cite{bib:cwallskyO3aPowerFlux}, we use the full O3 data set and exploit loose coherence in the first stage while combining SFT coefficients coherently from the two interferometers. The improved
strain sensitivity comes at an increased computational cost, however, for a fixed frequency range.
Restricting the search frequency band to 30--150 Hz makes the cost of this search acceptable. 

\subsection{Upper Limits}
We use a universal statistic~\cite{bib:universalstatistic} to calculate upper limits from the first search stage, as in the O3a \pf/ search~\cite{bib:cwallskyO3aPowerFlux}. Owing to origin in an Markov inequality that is agnostic to the underlying probability distribution, our upper limits are valid even in the presence of non-Gaussian noise and give nearly optimum upper limits when the noise is indeed Gaussian~\cite{bib:universalstatistic}. The calculated upper limits are based on the overall noise level in each demodulated 1/16 Hz search band and the largest power value within that band~\cite{bib:universalstatistic}. We calculate 95\% upper limits for circular polarization corresponding to the best-case scenario, and linear polarization corresponding to worst-case scenario. Additionally, we provide estimated upper limits for a population averaged over random sky locations and polarizations.

\begin{figure*}[htbp]
  \begin{center}
    \ifshowfigs
    \includegraphics[width=6.5in]{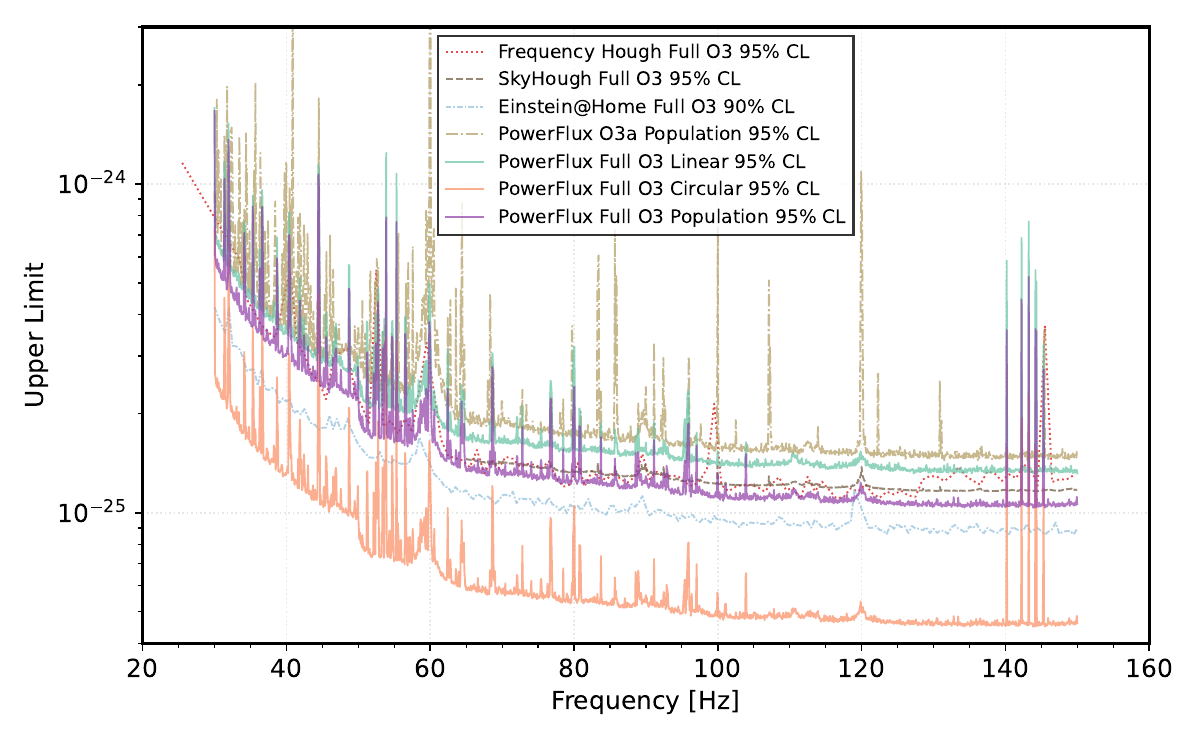}
    \fi
    \caption{Upper limits on gravitational strain amplitude for this analysis compared to other O3 analysis including the \pf\ O3a analysis.
      The green curve shows worst-case (linearly polarized)
      $95$\%~CL upper limits in each 1/16 Hz sub-band. The lowest orange curve shows the best-case (circularly polarized) upper limits. The middle (purple) curve shows approximate
      population-averaged all-sky upper limits inferred from the circularly polarized limits. As noted in \sect{results_upper_limits}, these pipelines cover different parameter space volumes, and hence care must be used when comparing the results. Additionally, the Einstein@Home search sets 90\% confidence upper limits unlike \pf\ and \sh, which use 95\% intervals.}
    \label{fig:upper_limits}
  \end{center}
\end{figure*}

\begin{figure*}[htbp]
  \begin{center}
    \ifshowfigs
    \includegraphics[width=6.5in]{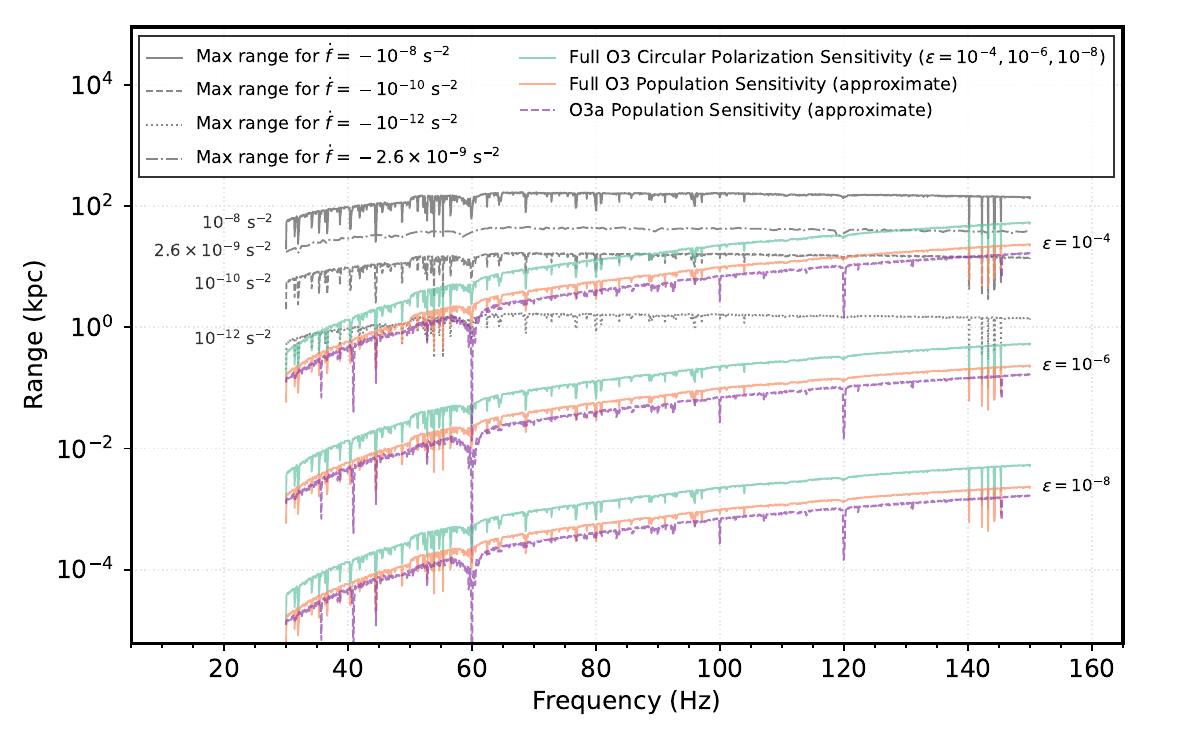}
    \fi
    \caption{Ranges (kpc) of this search as recast from the calculated upper limits on gravitational strain amplitudes. The green, orange, and purple curves show ranges for full-O3 circular-polarization, full-O3 population-averaged, and O3a population-averaged results for assumed ellipticities of $\epsilon = 10^{-4}, 10^{-6}$, and $10^{-8}$. The four gray curves show maximum ranges possible for maximum spindown magnitudes of $10^{-8}$, 2.6$\times$10$^{-9}$ (Einstein@Home search maximum), $10^{-10}$, and $10^{-12}$ Hz/s.
      }
    \label{fig:range}
  \end{center}
\end{figure*}

The upper limits are naturally not valid for the cleaned regions shown in \fig{excluded_bins}. We use software injections to validate our upper limits in the retained subbands, as illustrated in \fig{ul_vs_strain}. We exclude vetoed frequency bins from all software injections. The plots show calculated linear-polarization upper limits as a function of the injected strains. Since our upper limits are 95\% frequentist, we expect the points to be above the diagonal $> 95\%$ of the times. The flat regions on the left correspond to low strain values where we do not have enough sensitivity to produce useful upper limits.

\subsection{Outlier Followup}
\label{sec:outlier_followup}
Outliers produced from the first stage are analyzed by subsequent stages of loose coherence, with increasing constraints on the effective coherence time. Steadily decreasing values of the loose coherence phase-mismatch $\delta$ demands increasing temporal coherence with each iteration and hence, help distinguish between signals that could potentially be astrophysical in source from instrumental artifacts~\cite{bib:loosecoherence,bib:cwallskyS5}. Additionally, we step more finely in frequency, spindown, and sky points with each successive stage, which improves parameter reconstruction resolution.

Outliers obtained from stage 0 with SNR $> 7$ are clustered  according to their frequency, spindown, and sky locations. The clustering algorithm we use works as follows: We start by sorting all the outliers in descending order of their SNRs. We then take the loudest outlier in the list and select all other outliers within a certain pre-defined distance. We assign all these outliers into a single cluster. We then remove all the outliers just assigned to a cluster from the list and select the next loudest outlier. We repeat this until all outliers have been assigned to their respective clusters. The distance measure is based on the frequency, spindown, and sky location and the thresholds are determined with injections. Once the clustering is complete, we take the loudest 10 outliers within each cluster and follow up on them with stage 1. After each stage $> 0$, we require the SNR to exceed a threshold empirically determined from software injections and require the reconstructed parameters to be consistent with those of the previous stage. \tab{followup_parameters} shows the SNR increase constraints that were required, along with the search parameters for each stage.

We use software injections to determine suitable search parameters as well as to validate the followup process~\cite{bib:cwallskyS5}. In addition to the above mentioned criteria between stages, an injection is considered recovered only if its inferred parameters lie within 2.5 mHz in frequency, $\sci{3}{-10}~\mathrm{Hz}/\mathrm{s}$ in spindown, and $28.5~\mathrm{rad}~\mathrm{Hz} / f$ in sky location from the true injected values. \fig{injection_recovery} shows injection recovery efficiencies for the major 30 Hz sub-bands (30-60 Hz, 60-90 Hz, 90-120 Hz, 120-150 Hz). Recovery for the three \pf\ stages are shown separately. The vertical dashed lines represent a relative strain, the ratio of injected strain to upper limit on that band, equal to 1, and the horizontal dashed lines represent 95\% recovery efficiency.

\begin{table*}[htbp]
\begin{center}
\begin{tabular}{rrllllllcc}
\hline
Label & \multicolumn{1}{c}{Frequency} & \multicolumn{1}{c}{Spindown} & \multicolumn{1}{c}{$\RAJ$} & \multicolumn{1}{c}{$\DECJ$} & \multicolumn{1}{c}{$h_0$} & \multicolumn{1}{c}{UL} & \multicolumn{1}{c}{UL} & \multicolumn{1}{c}{Detected?}& \multicolumn{1}{c}{Bayes}\\
 & \multicolumn{1}{c}{Hz} & \multicolumn{1}{c}{nHz/s} & \multicolumn{1}{c}{degrees} & \multicolumn{1}{c}{degrees} & \multicolumn{1}{c}{true} & \multicolumn{1}{c}{sig bin}  & \multicolumn{1}{c}{ctrl bins} & & \multicolumn{1}{c}{Factor}\\
\hline \hline
Inj3   &  108.857159  & $\sci{-1.46}{-8}$   &  178.37257     &  -33.43660  & $\sci{1.30}{-25}$ & $\sci{1.4}{-25}$ & $\sci{1.4}{-25}$ & Yes & 482\\
Inj5   & 52.808324  & $\sci{-4.03}{-9}$     &  302.62664     &  -83.83914 & $\sci{3.99}{-25}$ & $\sci{4.8}{-25}$ & $\sci{2.8}{-25}$ & {Yes}  & 4542\\
Inj6   &  145.338566 & $\sci{-6.73}{0\mathrm{\;\;\;}}$     &  358.75095     &  -65.42262 & $\sci{3.84}{-25}$ & $\sci{3.3}{-25}$ & $\sci{1.4}{-25}$ & {Yes} & 6640\\
Inj11  &   31.424696 & $\sci{-5.07}{-4}$    &  285.09733     &  -58.27209 & $\sci{3.17}{-25}$ & $\sci{1.2}{-24}$ & $\sci{7.3}{-25}$ & No & 16\\
Inj12  &   37.706390 & $\sci{-6.25}{0\mathrm{\;\;\;}}$     &  331.85267     &  -16.97288 & $\sci{2.63}{-25}$ & $\sci{3.8}{-25}$ & $\sci{3.9}{-25}$ & Yes & 261\\
\hline
\end{tabular}\\
\caption[Parameters of hardware injections]{
  Parameters of hardware-injected simulated isolated-source continuous wave signals during the O3 run with frequencies falling within the analyzed 30-150 Hz range (epoch GPS 1253977218), along with upper limits in the nominal signal bins and averaged over the six nearest control bins. In the previous O3a \pf\ search, only injections Inj5 and Inj6 were detected. In a previous full-O3 search using the \fh\ and \sh\ pipelines, only injections Inj3, Inj5, and Inj6 were detected. The successful recovery here of Inj12 confirms the improvement in sensitivity expected with the search method presented here.}
\label{tab:hwinjections}
\end{center}
\end{table*}

\def\mc#1{\multicolumn{1}{c}{#1}}
\begin{table*}[htb]
\begin{center}
  \begin{tabular}{lrrrr}\hline
    \T\B                                       &    \mc{30-60  Hz}        &  \mc{60-90 Hz}         & \mc{90-120 Hz}           & \mc{120-150 Hz} \\
    \hline\hline
    \T\B Stage 0                                     &  317,002 & 1,119,150 & 567,252 & 853,700 \\
    \hline
    \T\B Stage 1                                     &  18,732 & 8,444 & 1,049 & 1,158  \\
    \hline
    \T\B Stage 2                                     &  8,606 & 2,614 & 213 & 144   \\
    \hline\hline
    \T\B \pyfstat\ followup                         &  8,606 & 2,614 & 213 & 144   \\
    \hline
    \hline
    \T\B Bayes Factor $\ge 15$                        & 86 & 14 & 6 & 6 \\
    \T\B Hardware Injections                        & 12 & 0 & 6 & 6 \\
    \T\B Visible Artifacts                          & 64 & 14 & 0 & 0 \\
    \hline
  \end{tabular}
  \caption{Counts of outliers surviving different stages of the search.
    Survivors from stage-2 followup are first manually examined for contamination from instrumental artifacts through the use of strain histograms and then followed up with \pyfstat\ as described in \sect{outlier_followup}.}
\label{tab:outlier_counts}
\end{center}
\end{table*}

After three stages of \pf, we manually inspect the outliers to look for contamination via strain histograms~\cite{bib:StrainHistogram}, as
was done for the O3a search~\cite{bib:cwallskyO3aPowerFlux}. Ruling out clearly contaminated signals, we then use Markov Chain Monte Carlo (MCMC) with the \pyfstat~\cite{bib:pyFstat} pipeline to follow up on the final list of outliers.

The \pyfstat\ pipeline uses semi-coherent summing of the \Fstat~\cite{bib:JKS} with Markov Chain Monte Carlo sampling of the posterior to give a detection statistic as well as to produce a posterior for the reconstructed parameters~\cite{bib:AshtonPrix}. We follow an implementation described in~\cite{bib:TenorioEtal} where it was used to follow up on outliers from the O2 run. In particular, we create a ladder of steadily decreasing numbers of segments of increasing duration with which to subdivide the full O3 run and run multiple stages where the posterior from one stage is then used to shrink the prior for the next stage~\cite{bib:TenorioEtal}. We use a random sample of 600 sky locations away with the same declination as the candidate source, but separated by more than 90 degrees in right ascension, in order to calculate a background distribution for each specific frequency band~\cite{bib:TenorioEtal}. We then use this background distribution along with the change in the \Fstat\ value in the last stage of the search to compute a Bayes factor by comparing the probability distribution of having a signal in the given data with the probability distribution of its being due to a noise background~\cite{bib:TenorioEtal}. We use software injections to calibrate the appropriate Bayes factor values to set a cutoff for what value would constitute large enough for an outlier to potentially be from a true signal. We obtain a threshold of 100, so any outlier with Bayes factor less than that can be safely discarded.

\section{Results}
\label{sec:results}

The first stage of the \pf\ search leads to upper limits and a list of outliers to follow up. The upper limit on a given band is valid regardless of whether or not it produces outliers~\cite{bib:cwallskyS6,bib:cwallskyO3aPowerFlux}. Additionally, the upper limits are robust with respect to non-Gaussian noise, owing to the use of the universal statistic~\cite{bib:universalstatistic}.

\sect{results_upper_limits} and \sect{results_outliers},\ respectively discuss the upper limits and outliers we obtain. In \sect{hardware_injections}, we present results from the recovery of hardware injections.

\subsection{Upper Limits}
\label{sec:results_upper_limits}

\fig{upper_limits} shows a graph of upper limits as a function of the search frequency. The green curve shows the linear-polarization or worst-case upper limits, while the orange curve shows the circular-polarization or best-case upper limits. The dashed and dotted curves show O3 results from other search pipelines, including the O3a \pf\ search~\cite{bib:cwallskyO3aPowerFlux}, full-O3 \fh\, and \sh\ searches~\cite{bib:cwallskyO3FourPipelines}, and full-O3 Einstein@Home distributed-computing
search~\cite{bib:EatHO3}. It is important, however, to use care in these comparisons as the searches cover different parameter spaces. For example, the Einstein@Home search covers only about 26\% of the spindown range of this search. Results from similar but less sensitive searches of the O1 and O2 data sets are not included.\footnote{The Falcon
pipeline~\cite{bib:Falconpaper,bib:cwallskyFalconO2LowFreq,bib:cwallskyFalconO3aMidFreq}, which pioneered the use of loose coherence in the first stage of a hierarchical analysis,
carried out a search of the O2 data in this frequency band~\cite{bib:cwallskyFalconO2LowFreq}
and obtained comparable strain sensitivity, but for only a tiny allowed
spindown range ($<\,$$10^{-12}$ Hz/s), limiting astrophysical range.}

To enable a straightforward comparison to the other pipelines, we also present estimated population-averaged upper limits~\cite{bib:cwallskyO3aPowerFlux}. This estimate is produced by multiplying the circular upper limits by a factor of 2.3, determined empirically for control bands using an ensemble of random sky locations and random stellar orientations~\cite{bib:cwallskyO3aPowerFlux}.

Each upper limit displayed is maximized over the entire sky except for a small region near the ecliptic poles~\cite{bib:cwallskyS5,bib:cwallskyO1paper1}. This region is excluded as it is prone to a large number of detector artifacts owing to stationary frequency evolution from the combination of frequency derivative and Doppler modulation~\cite{bib:cwallskyS5,bib:cwallskyO1paper1}.

Our results improve upon the earlier \pf\ O3a results by a median factor of $\sim 1.4$. Improvements with respect to the \fh\ and \sh\ full-O3 searches are frequency-dependent, with median factors of $\sim 1.1$ for both. In contrast, the full-O3 Einstein@Home distributed-computing search~\cite{bib:EatHO3}, which covers the 20-800 Hz band improves
upon these results by a median factor of $\sim1.2$ in the common range of frequency overlap and spindown overlap (26\%\ of the spindown range probed here).

\begin{table}[htb]
  \begin{center}
    \input{outliers.tex}
    \caption{Parameters of the outliers surviving \pyfstat\ followup with Bayes Factor $\ge 15$.
      Each outlier listed is the loudest after clustering in frequency, spin-down, and sky location. Outliers marked with an * are from hardware injections. A list of all hardware injections within the paramater space of this search can be found in \tab{hwinjections}.}
\label{tab:outliers}
\end{center}
\end{table}

The improvements compared to the O3a \pf\ analysis can be attributed to a number of reasons. First, this analysis uses the full O3 data set which is nearly twice as long as the O3a data set~\cite{bib:O3OpenData}. Second, unlike in the O3a search, we use loose coherence in the initial stage with a $\delta$ parameter value of $\pi/2$, which improves sensitivity via longer effective coherence time, at the expense of higher computational cost~\cite{bib:loosecoherence,bib:cwallskyS5}. Third, we combine the interferometer SFTs coherently in the initial stage~\cite{bib:cwallskyO1paper1}.

As a large number of contaminating line artifacts were present, particularly in H1 data, we clean the SFTs by replacing polluted bands with randomly generated Gaussian noise~\cite{bib:lal}. This method naturally means that upper limits calculated for those bands would not be valid, so those bands are excluded from the results presented here. \fig{excluded_bins} shows the bands that were cleaned out.

The upper limits on source strain amplitudes shown in \fig{upper_limits} can be recast into lower limits on distance ranges at which source neutron stars with certain assumed ellipticities can reside~\cite{bib:LRRReview,bib:cwallskyS4}. \fig{range} shows these implied lower limits on ranges (kpc) for assumed ellipticity values of $\epsilon = 10^{-4}, 10^{-6}$, and $10^{-8}$. The ranges are shown for the best-case circular popularization as well as population-averaged ranges. We also show our search's implied maximum ranges as a function of the frequency for a gravitar with maximum spindown magnitudes of $10^{-8}, 10^{-10}$, and $10^{-12}$ Hz/s.

\subsection{Hardware Injections}
\label{sec:hardware_injections}

A total of 18 hardware injections-- 16 isolated and two binary sources-- were added to the data in the O3 run~\cite{bib:hwinjectionpaper}. Out of the 16 isolated-star injections, five injections fall within the frequency range we analyze~\cite{bib:hwinjectionpaper}. We successfully recover four out of the five hardware injections. \tab{hwinjections} shows the list of relevant hardware injections, their parameters including injected strains, corresponding upper limits produced and the final Bayes factors we obtain. The control upper limits show 95\% upper limits for the nearest six neighboring sub-bands to give a rough estimate of expected upper limit that would have been obtained in the absence of a signal. For reference,
only two of the five injections were successfully recovered in the O3a \pf\ search~\cite{bib:cwallskyO3aPowerFlux}, and only three of the five were recovered
in a previous full-O3 search~\cite{bib:cwallskyO3FourPipelines} using other pipelines, consistent with the improved sensitivity of the
method presented here.

\subsection{Outliers}
\label{sec:results_outliers}

In addition to producing upper limits, stage 0 produces an initial list of outliers, corresponding to templates with excess power. We follow up on these outliers via subsequent stages, in which we impose progressively tighter constraints. In the absence of a true signal or of a hardware
injection, we would ideally expect the number of outliers to decrease in subsequent stages because of these tighter constraints. In practice, particularly severe detector artifacts can also persist.
\tab{outlier_counts} shows the number of outliers we obtain for various stages of the search.

For the outliers that survive stage 2, we use strain histograms in which we plot the expected spectral shape overlaid upon the actual amplitude spectral density (ASD) to check if a particular band is clearly affected by an artifact. After the \pf\ stages, we use Markov Chain Monte Carlo (MCMC) followup with the use of the \pyfstat\ pipeline to obtain a Bayes factor for each of the surviving outliers. We use injections to determine appropriate Bayes factors as described in \sect{outlier_followup}. \tab{outliers} shows the Bayes factors $\ge15$ for the final list of clusters of outliers. As can be seen, all outliers with appreciable Bayes factors have origins in hardware injections. The largest Bayes factor for an outlier not coming from a hardware injection and not having a visible artifact is 27, well below the threshold of 100 used to identify a plausible signal.

\section{Conclusions}
\label{sec:conclusions}

We have performed a deeper search for continuous gravitational waves from isolated neutrons stars, building upon a previous analysis~\cite{bib:cwallskyO3aPowerFlux}. Limiting the frequency band of the search allowed us to go deeper through the use of the full O3 data set, loose coherence, and coherent summing of SFTs across detectors in the initial stage. Additionally, line cleaning the SFTs before performing the analysis made the number of outliers to follow up on much more manageable.

While we failed to detect a credible signal as all surviving outliers have origins in either hardware injections or instrumental artifacts, we do set upper limits on the strain amplitudes as shown in \fig{upper_limits}.

The best upper limit obtained for circular polarization is $\sim\sci{4.5}{-26}$ near 144 Hz, and for linearly polarization is $\sim \sci{1.3}{-25}$. The best estimated population-averaged upper limit is $\sim\sci{1.0}{-25}$, These limits improve on our O3a search by a median factor of $\sim 1.4$. Improvements with respect to the \fh\ and \sh\ full-O3 searches~\cite{bib:cwallskyO3FourPipelines} are frequency-dependent, with median factors of $\sim 1.1$ for both. Our results are less constraining by a median factor of $\sim 1.2$
than the full-O3 Einstein@Home search, in the overlapping frequency band and spindown range, but cover a spindown range four times larger and hence probe to greater distances
in the galaxy in the band searched.

The fourth LIGO-Virgo-KAGRA observation run O4 started May 24, 2023 with a planned duration of $\sim$20 months.
The improved detector noise in the O4 data and longer observation period offer the prospect of improved
strain sensitivity and, ideally, a continuous wave signal detection at last.

\section{Acknowledgements}

We gratefully acknowledge useful discussions and long collaboration with current and former
colleagues in the LIGO-Virgo-KAGRA continuous waves working group.
For these results, in particular, we thank Evan Goetz, Ansel Neunzert, and Grant Weldon for
spectral line investigations of the O3 data, Rodrigo Tenorio and David Keitel for creating the MCMC
\pyfstat\ pipeline, and Vladimir Dergachev for creating the \pf\ pipeline. We also thank the two anonymous journal referees for constructive comments and suggestions. This work was supported in part by National Science Foundation Award No. PHY-2110181.

This research has made use of data or software obtained from the Gravitational Wave Open Science Center (gwosc.org), a service of the LIGO Scientific Collaboration, the Virgo Collaboration, and KAGRA. This material is based upon work supported by NSF's LIGO Laboratory which is a major facility fully funded by the National Science Foundation, as well as the Science and Technology Facilities Council (STFC) of the United Kingdom, the Max-Planck-Society (MPS), and the State of Niedersachsen/Germany for support of the construction of Advanced LIGO and construction and operation of the GEO600 detector. Additional support for Advanced LIGO was provided by the Australian Research Council. Virgo is funded, through the European Gravitational Observatory (EGO), by the French Centre National de Recherche Scientifique (CNRS), the Italian Istituto Nazionale di Fisica Nucleare (INFN) and the Dutch Nikhef, with contributions by institutions from Belgium, Germany, Greece, Hungary, Ireland, Japan, Monaco, Poland, Portugal, and Spain. KAGRA is supported by Ministry of Education, Culture, Sports, Science and Technology (MEXT), Japan Society for the Promotion of Science (JSPS) in Japan; National Research Foundation (NRF) and Ministry of Science and ICT (MSIT) in Korea; Academia Sinica (AS) and National Science and Technology Council (NSTC) in Taiwan.

The authors are grateful for computational resources provided by LIGO Laboratory and supported by National Science Foundation Grants PHY-0757058 and PHY-0823459.

This research was done in part using services provided by the OSG Consortium \cite{bib:osg1,bib:osg2,bib:osg3,bib:osg4}, which is supported by the National Science Foundation Award Nos. 2030508 and 1836650.

This research was enabled in part by support provided by the Simon Fraser University Cedar Cluster (\href{https://www.sfu.ca/research/supercomputer-cedar}{sfu.ca/research/supercomputer-cedar}) and the Digital Research Alliance of Canada (\href{https://www.alliancecan.ca/}{alliancecan.ca}).

\bibliography{probing-more-deeply-full-o3}

\let\author\myauthor
\let\affiliation\myaffiliation
\let\maketitle\mymaketitle

\end{document}

%% file: outliers.tex
\def\mc#1{\multicolumn{1}{c}{#1}}
\begin{tabular}{rrrrr}\hline
\mc{$f$} & \mc{$df/dt$} & \mc{R.A.} & \mc{Dec.} & \mc{Bayes} \\
\mc{(Hz)} & \mc{(nHz/s)} & \mc{(radians)} & \mc{(radians)} & \mc{Factor} \\ 
\hline
\hline
31.4437 & -3.651 & 0.509 & 0.386 & 16
\\
31.9900 & -1.035 & 5.237 & -1.037 & 20
\\
37.7064* & -6.250 & 5.792 & -0.299 & 261*
\\
38.6659 & -0.265 & 2.605 & -1.414 & 27
\\
38.7084 & 0.009 & 4.671 & -0.839 & 18
\\
38.7121 & 0.090 & 0.532 & 1.483 & 21
\\
38.7123 & -0.286 & 4.729 & -1.003 & 16
\\
40.4638 & -3.268 & 2.832 & -0.579 & 19
\\
40.4934 & -6.005 & 0.014 & 0.435 & 15
\\
40.4994 & -5.265 & 4.357 & 0.726 & 18
\\
40.5085 & -7.749 & 4.033 & -0.613 & 20
\\
41.6605 & 0.205 & 4.770 & 1.539 & 25
\\
44.4287 & -5.088 & 3.684 & -0.438 & 15
\\
44.5042 & -0.870 & 5.942 & -0.303 & 18
\\
44.5237 & -4.166 & 3.689 & -0.394 & 16
\\
44.5410 & -7.255 & 4.892 & -0.799 & 16
\\
44.5484 & -6.860 & 2.802 & 0.294 & 15
\\
44.5604 & -9.295 & 2.029 & 0.427 & 18
\\
44.5668 & -8.580 & 3.090 & 0.665 & 20
\\
44.5668 & -8.051 & 6.192 & -0.914 & 18
\\
44.5889 & -7.549 & 1.939 & 0.655 & 18
\\
44.5992 & -9.492 & 2.774 & 0.558 & 19
\\
44.6078 & -9.709 & 4.532 & -0.584 & 16
\\
45.5851 & -0.088 & 5.544 & -0.479 & 17
\\
46.1913 & 0.560 & 6.239 & 0.953 & 22
\\
46.1941 & -0.270 & 2.352 & -1.507 & 22
\\
49.9925 & 0.331 & 0.084 & 1.500 & 23
\\
49.9971 & -0.361 & 3.423 & -1.487 & 27
\\
52.8083* & 0.000 & 5.276 & -1.463 & 4542*
\\
53.3882 & 0.965 & 2.395 & 1.143 & 18
\\
53.4048 & 0.043 & 1.804 & 0.709 & 15
\\
53.6997 & -0.916 & 5.927 & -0.580 & 15
\\
55.5455 & -0.444 & 4.769 & -1.415 & 17
\\
56.5645 & 0.513 & 5.975 & 0.874 & 17
\\
59.4917 & -1.315 & 0.437 & -0.010 & 61
\\
80.0337 & -2.555 & 1.405 & -0.626 & 21
\\
80.0338 & -2.486 & 1.270 & -0.722 & 33
\\
83.3048 & -1.446 & 4.177 & -0.828 & 22
\\
85.7160 & 0.990 & 0.117 & -0.503 & 21
\\
108.8572* & 0.000 & 3.114 & -0.581 & 482*
\\
145.3386* & -6.730 & 6.262 & -1.142 & 6640*
\\
\hline
\end{tabular}